# High temperature superconductivity: Cooper pairs in trap

Wei Ruan

**Affiliations:**

Key Laboratory of Inorganic Functional Materials and Devices, Shanghai Institute of Ceramics, Chinese Academy of Sciences, Shanghai 200050, China

Email: ruanwei@mail.sic.ac.cn   or   revenger@shu.edu.cn

**Abstract**

The tremendous efforts to unveil high temperature superconductivity (HTSC) have been devoted to the search of the mechanism underlying Cooper pairs which, however, remains a mysterious subject of vigorous debate, let alone many other mysteries like the pseudogap state, the peculiar Homes' law, the unneglibible electron-phonon interaction, the stripe, the universal nodal Fermi velocity, etc. Most of subsequent works either bring in more controversies or swell the list of mysteries. Here I tentatively propose a whole new perspective on the basis of low-dimensional Bose-Einstein Condensation (LDBEC), which possibly makes lots of those mysteries correlated and understood in single picture.

It is believed that we know as much about the cuprate HTSC as we don't, due to so many controversies and mysteries. To find out what exactly underlies the cuprate HTSC, we first requires an in-depth understanding of the cuprate pairing mechanism. Anderson's 'resonating valence bond' (RVB) theory expresses a magnetic pairing idea — the strong Coulomb repulsion in these two-dimensional (2D) quantum spin-1/2 systems yields an antiferromagnetic (AF) Mott insulator where the singlet electron pairs naturally form arising from the AF superexchange interaction [1,2]. Recent works further discussed and reviewed the magnetic pairing mechanism to elucidate the cuprate HTSC — superconductivity without phonon [3,4]. No matter in what specific form or model, the magnetically mediated pairing mechanism has provided so many significant insights and guidances in pursuit of the essence of the cuprate HTSC. However, recent rise of experimental evidences aroused people's interest in the electron-phonon interaction (EPI) [5-8], indicating the unnegligible role of EPI that has been almost oblivious in HTSC but dominant in pairing in the Bardeen-Cooper-Schrieffer (BCS) theory.

Anyway, it is believed that the more we know about the pairing mechanism, the closer we are to unravel the mystery of HTSC. However, more and more discoveries reveal that the problem is far more complicated: it is the ability of the superconductor to carry a supercurrent that has been destroyed, namely that the phase stiffness is destroyed, while the Cooper pairs continue to exist above $T_c$ — the preformed Cooper pairs [1,9-12]. Furthermore, there are still many mysteries that seem unable to be accounted for in terms of Cooper pairs, i.e., the Homes' law [13,14], stripes [1,15], the peculiar metallic normal state [16], etc. It seems that: while we are focusing too much on the Cooper pairs that is taken for granted to be responsible for the 'crime' (superconductivity), the other main 'culprits' have escaped away.

From the fundamental perspective of quantum physics, what could be responsible for HTSC? In 1938, Fritz London proposed Bose-Einstein condensation (BEC) as a mechanism for the superfluidity in liquid helium, which led to the possible interpretation of superconductivity as a superfluidity of charged particles in terms of BEC (Fig. 1a). Chen *et al*. have reviewed and discussed the connection between BEC and HTSC [17], and there are also other works involving the concept of BEC that have discussed HTSC with different theoretical models. However, those arguments still remain highly contentious, and many of them are often different or even incompatible with each other. Most importantly, how a specific theory based on BEC can account for so many mysteries and crucial problems in high-$T_c$ cuprates is still elusive or unknown, or in other words, so far there is no such a theory.

In this paper, without any complicated theoretical calculations or models, I propose a tentative perspective that the cuprate HTSC could probably be regarded as the manifestation of LDBEC reminiscent of the trapped low-dimensional bose gas, which is fundamentally different from those in any previous works involving the concept of BEC and could yield a simple and clear phenomenological picture, supported by plenty of previous evidences on high-$T_c$ cuprates, that possibly accounts for lots of mysteries in the cuprate HTSC.

According to the Bose-Einstein distribution, the total number of particles $N$ in the bose system is given by:

$$N = \sum_k \frac{1}{\exp[\beta(\varepsilon_k - \mu) - 1]} \quad (1)$$

and the particle density $n$ is given by:

$$n = \int_0^\infty f(\varepsilon)\rho(\varepsilon)d\varepsilon, \quad f(\varepsilon)\rho(\varepsilon) = \frac{\rho(\varepsilon)}{\exp(\frac{\varepsilon-\mu}{k_B T})-1} \quad (2)$$

where $\beta = (k_B T)^{-1}$, the $f(\varepsilon)$ is the Bose-Einstein distribution function, $\rho(\varepsilon)$ the density of states (DOS). Einstein realized that the number of boson particles in the lowest energy quantum state or ground state becomes infinite as the chemical potential $\mu$ increases and reaches zero with decreasing temperature, more precisely we can say that the occupation number of the lowest energy quantum state becomes macroscopic when the chemical potential $\mu$ becomes equal to the energy $\varepsilon$ of the lowest quantum state — Bose-Einstein Condensation (BEC) [18]. From the point of view of DOS which is $\rho(\varepsilon) \propto \varepsilon^{(D/2-1)}$, where $D$ is the dimension of the system, since the DOS $\rho(\varepsilon)$ decreases with decreasing $\varepsilon$ or temperature in 3D case (Fig. 1b), at sufficiently low temperature it becomes impossible to thermally occupy the low energy state while keeping a constant chemical potential or density, and thus a macroscopic number of bosons crowd into the ground state, which is one of the reasons for the absence of BEC in 1D or 2D case where the temperature-dependent DOS behaves differently [19]. Essentially, the BEC is actually a statistical effect or phase transition driven by the quantum statistical mechanics, which can be mediated by decreasing temperature to significantly low value that significantly decreases the DOS and shrinks the phase space available to the bosons.

Normally, how many ways are there to achieve BEC? The first one could be increasing the density of bosons, as exemplified by Demokritov *et al*'s recent work where they increase the density of bosons to achieve the room-temperature BEC in a gas of magnons by microwave pumping [20]; The second one could be the most typical way as decreasing temperature to sufficiently low value, exemplified by conventional superconductor if we regard it as a special case where the condensation and pair formation temperatures coincide [17], or the condensation occurs as soon as the electrons of opposing spin and momentum are bound into pairs of same total momentum, which is rather a condensation in velocity or momentum space that was first recognized by Fritz London, unlike the well-known condensation in coordinate or real space as the gas-liquid phase transition.

Then, even if we know exactly how the Cooper pairs (bosons) form, how could BEC occur in the high-$T_c$ cuprate superconductors with such high temperature and low carrier density? It should be noted that being boson is not equivalent to exihibiting BEC (Fig. 1). What's even worse is the problem associated with the correlation property that arises from low dimensionality: a true or long-range ordered BEC could never occur in 1D or 2D case at non-zero temperature due to the thermal phase fluctuations that distablize the condensate [18,19,21-23], instead, the systems are characterized by divided blocks with true condensate in each block but no correlation with each other, namely a 'quasicondensate' (Fig. 2a) [19,21]. Table 1 gives a more clear view of these serious problems that the main features of high-$T_c$ cuprates are detrimental to achieving a true BEC. One possible way out of this predicament could be introducing a strong attraction as many theoretical works did [17,24], however, what exactly binds the holes into Cooper pairs in cuprates is still under debate, let alone that it even remains controversial whether a real attraction to bind holes exists in cuprates [25]. Then, what is exactly happening in the 2D Cu-O superconducting layers at such high temperature?

To tackle these problems, it's worth noting the theoretical works on the trapped bose gas. A trapping potential, i.e., harmonic potential with trapping frequency $\omega$, could constrain the phase space between energy $\varepsilon$ and $\varepsilon + d\varepsilon$ and thus decrease the volume and DOS available to the bose system [26], making the chemical potential $\mu$, DOS $\rho(\varepsilon)$ and even $T_c$ strongly dependent on the trapping potential [18,19,21]:

$$\mu \propto \hbar\omega \ , \quad \rho(\varepsilon) \propto \frac{\varepsilon^{(D-1)}}{(\hbar\omega)^D} \ , \quad T_c \propto \hbar\omega \qquad (3)$$

(where the dimension $D = 1$ or 2, and other involved parameters like number of particles $N$, etc. are not shown in above equation), causing BEC not only in momentum space as in superfluid helium, but also in coordinate space, which is quite a peculiar feature of trapped bose gas [18]. Furthermore, the phase fluctuations could be quenched because the trapping potential significantly modifies the DOS behavior, leading to the true BEC in low dimension [18,19]. As a result, the occurrence of BEC even in low dimension doesn't have to depend uniquely on sufficiently low temperature any more, as long as the trapping potential is strong enough, since it could shrink the phase space and strongly modify the DOS behavior instead of cooling. This gives quite important implication for HTSC, however, few works have ever linked the trapped bose gas with HTSC. So far it's indeed hard to say that the high-$T_c$ superconductor behaves exactly as the trapped bose gas does, however, it's tempting to presume that the high-$T_c$ superconductors may share some basic essentials of trapped bose gas — something behaves as the trapping potential shrinking the phase space available to the preformed Cooper pairs instead of cooling, leading to a high temperature BEC even in low dimension.

There are indeed evidences in support of the analogy between HTSC and trapped bose gas. Figure 2b shows the state diagram of the trapped low-dimensional bose gas [27], which is strikingly analogous to the familiar phase diagram of the hole-doped cuprate high-$T_c$ superconductors (Fig. 2c); Gomes *et al*. show the temperature evolution of the local superconducting gaps of $Bi_2Sr_2CaCu_2O_{8+\delta}$, indicating that the gaps exhibit a 'quasicondensate'-like behavior above $T_c$ and a 'true condensate'-like behavior below $T_c$ [10], which is analogous to the BEC behavior of trapped low-dimensional bose gas (Fig. 2b); The temperature-dependent intensity of the familiar '$\pi$–resonance' in high-$T_c$ cuprates shows a smeared out onset of transition as the doping level or hole concentration decreases [28], which is analogous to the BEC transition behavior of trapped low-dimensional bose gas as the boson number decreases [29]. More evidences could be found in support of such analogy, if we reexamine the numerous previous works on high-$T_c$ cuprates, implying that the cuprate HTSC could probably be regarded as the manifestation of LDBEC reminiscent of the trapped low-dimensional bose gas.

Therefore, it is important to know what the 'trapping potential' is in HTSC. I tentatively propose that the role of the 'trapping potential' could probably be played by the recently emerging phenomenon of EPI [5-8], which yields a dichotomy that naturally reconciles the controversy aroused by the EPI. Recent evidences unequivocally indicate the unnegligible role of EPI in high-$T_c$ cuprate superconductors [5-8], arousing the controversy against the magnetic pairing mechanism about what is responsible for the formation of Cooper pairs. However, the mechanism of how the EPI pairs the holes in high-$T_c$ cuprates seems so elusive and contentious, and what's most embarrassing are the unambiguous and solid evidences for both phenomena (EPI and spin-spin interaction) supporting that they are both strongly influencing the cuprate HTSC. The 'trapping potential' provides a natural way to overcome this dilemma: the holes pair up due to the spin-spin interaction or the magnetic pairing mechanism [1-5], and the EPI condenses the magnetically paired holes as a trapping potential. Furthermore, such dichotomy is supported by numerous previous evidences and mysteries:

(1) The scanning tunnelling microscopy (STM) results of Lee *et al.* show that a Cu-O bond-oriented electron-phonon interactions (EPI) is strongly influencing HTSC with a spatial anticorrelation between the local superconducting gaps $\Delta(r)$ and the phonon mode energy $\Omega(r)$ [8], which is consistent with lots of observations of phonon anomalies suggesting that the charge stripes along the Cu-O bonds are coupling with the particular phonon mode [5,6,30,31], and more importantly, the EPI in their STM results can not be simply related to pairing [8]. Note that Deutscher has suggested that there are two distinct energy scales in high-$T_c$ cuprates — one is associated with pairing, while the other one is associated with condensate of the paired charges [32].

(2) Many STM works have observed in the high-$T_c$ cuprates the Cu-O bond-oriented spatial modulations in local DOS or electronic structure with many features, i.e., direction (along the antinodal or Cu-O bond direction), periodicity, etc., quite similar to that of 'stripes' — the mysterious phenomenon that describes the 1D aggregation of holes into charge stripes separated by hole-poor AF insulating regions in the Cu-O planes of the cuprates (Fig. 3) [1,33-36]. Parker *et al.* further show the 'smoking gun' evidences: the Cu-O bond oriented modulations in local electronic state are due to the fluctuating stripes which are strongly influencing the pseudogap that: regions with stronger modulations exhibit larger pseudogaps or gaps in DOS at higher temperatures, see Fig. 4 in Ref. [37].

(3) One of the most mysterious phenomena in high-$T_c$ cuprates — pseudogap phase — is characterized by the opening of the gap in DOS near antinode (along the Cu-O bond direction) in the momentum space above $T_c$, leading to the breakup of the Fermi surface into disconnected segments — 'Fermi arcs' — that continuously shrink down to the nodal points (45° to the Cu-O bond direction) at $T_c$, with electronic states continuously consumed in the process [38,39]; In fact, lots of works using different techniques have proven that the pseudogap manifests itself as a gradual and continuous depletion of the normal-state DOS near Fermi level $E_F$ in the momentum space, which starts near $T^*$ (the pseudogap onset temperature) and continues to temperature below $T_c$ [11,40,41]. More intriguingly, previous works suggested that something else other than simply pairing, that relates to the gap in DOS, is hidden in the pseudogap phase at the antinode [42], and this hidden effect is suggested to be strongly coupled to the stripes [38,43-45]. Therefore, it is now known that there are at least two effects associated with the gap in DOS in the pseudogap phase [11,42] — one arising from pairing [9,11], and the other one unidentified but associated with the stripes [8,42]. So far, no works have been able to convincingly explain what exactly underlies this peculiar phenomenon in the peudogap phase, but what if it reflects the shrinkage of momentum space suggestive of the effect of the trapping potential, other than the preformed Cooper pairs?

(4) we can naturally get a LDBEC picture based on numerous previous evidences and the idea of 'trapped Cooper pairs' to describe the mysterious phenomena of HTSC (Fig. 4): around $T^*$, the aggregation of holes into stripes allows the formation of the fluctuating stripes array which favors the development of local AF correlations in the hole-free regions [15,46]; The stripes-related EPI, which behaves as a trapping potential enhancing the pseudogap or gap in DOS, starts and gradually develops with decreasing temperature [6,11,39-42,47]; As the temperature is cooled to $T_{pair}$ ($T_c < T_{pair} < T^*$) [11], pairing behavior of the holes emerges within the stripes, or the spin gap opens [1,9-12,46]; Meanwhile, the bosons (hole pairs) start to condense due to the shrinkage of momentum space induced by the 'trapping potential', in a gradual and continuous process of BEC with the long-range superconducting phase coherence gradually develops till $T_c$ [10]. Evidences can be easily found supporting almost every detail of this picture.

Finally, the dichotomy in the LDBEC picture could be compatible with most of the important and fundamental theories of high-$T_c$ cuprates, and seems to be able to make many mysteries compatible and correlated with each other in single picture, i.e., the unnegligible EPI, the spin-spin interaction, the preformed Cooper pairs, stripes, the phonon anomalies, the 'Fermi arcs', etc, and even possibly implies why the $T_c$ is high. However, such an idea of 'trapped Cooper pairs' is yet speculative with evidences, especially on EPI, obtained so far, here I propose it in hope of stimulating a whole new perspective to unravel the mystery of cuprate HTSC — Cooper pairs in trap.

**Acknowledgments:**

**Figure legends**

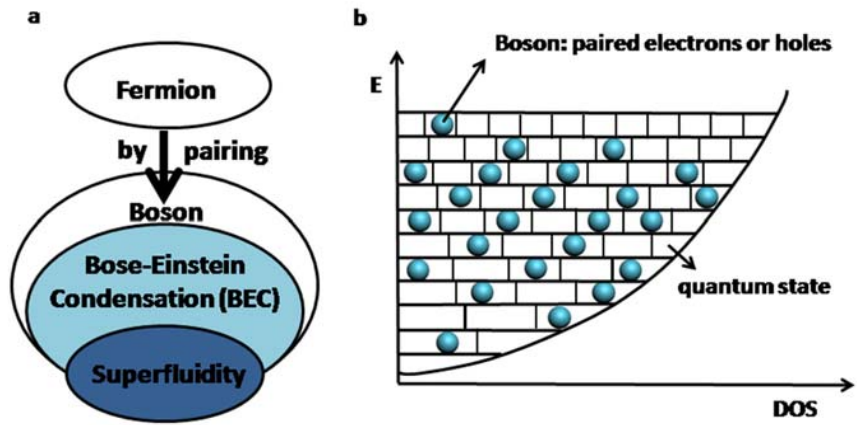

**Figure 1.** The textbook physics about BEC. **a**, the electrons (fermions with half-integer spin) can be bound into pairs (bosons with integer spin), i.e., by attraction arising from EPI at sufficiently low temperature in BCS theory; Owing to their bosonic nature that they are not subject to the Pauli exclusion principle, a large number of these bosons (electron pairs) can share the same quantum state in the process of BEC, and thus exhibit the phenomenon of frictionless or dissipationless flow of electrical current — superconductivity. **b**, The relationship of energy *E* versus DOS is exemplified by that in 3D systems.

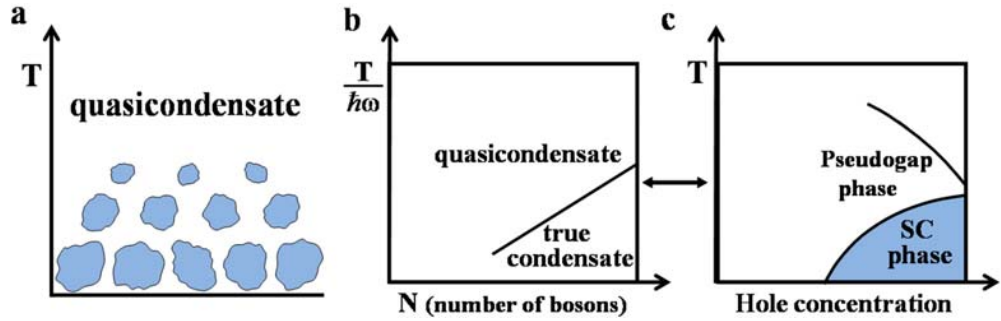

**Figure 2. a**, The schematic illustration of the 'quasicondensate' of the ideal low-dimensional bose system, see detailed descriptions in [19,21]. **b**, State diagram of the trapped 1D bose gas, the $T_c$ that divides the 'quasicondensate' and 'true condensate' is determined by the repulsive interaction coupling constant $g$, the trap frequency $\omega$, the number of bosons $N$, etc, see detailed descriptions in [27]. **c**, The well-known phase diagram of the hole-doped cuprate high-$T_c$ superconductors, the double-headed arrow between **b** and **c** indicates their analogy. (SC: superconducting)

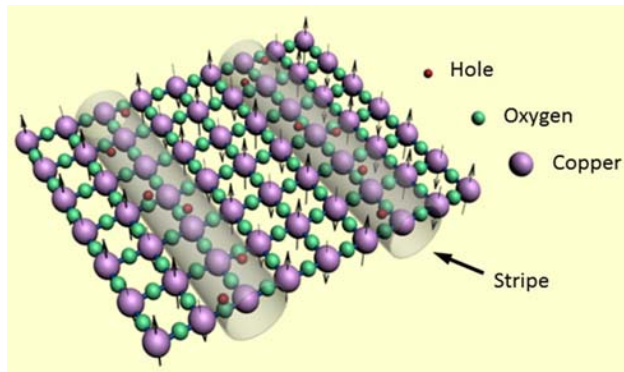

**Figure 3.** The schematic illustration of the stripes. The phenomenon of stripes describes the alternating stripes of holes and AF insulating regions in the Cu-O planes of the cuprates superconductors, which originates from the competition between the kinetic energy of holes and AF interaction energy, see the schematic illustration of the fluctuating stripes in Ref. [15].

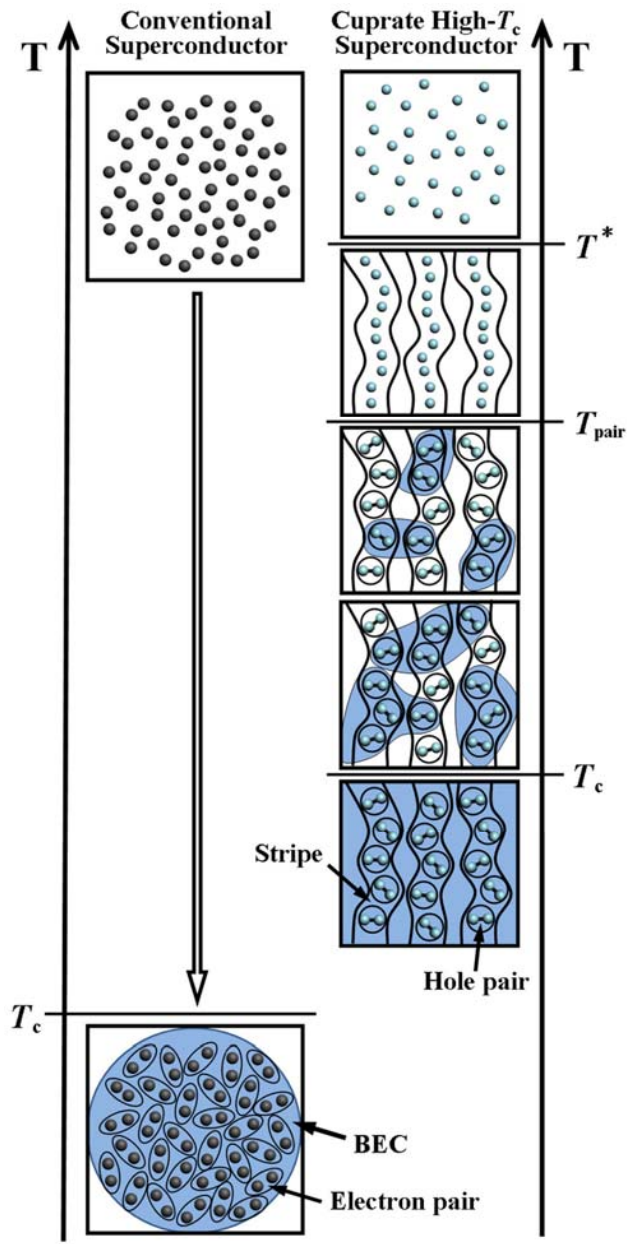

**Figure 4.** The schematic illustration of the temperature-dependent evolution of high-$T_c$ cuprate superconductors in comparison with that of conventional superconductors.

**Table 1.** Comparison of the superconducting parameters for two types of superconductors

|  | $T_c$ | Carrier density | Dimension |
|---|---|---|---|
| **Conventional superconductors** | Low (near 0 K) | $10^4$ nm$^{-3}$ (normal metal) | 3D |
| **High-$T_c$ cuprate superconductors** | High (up to around 160 K) | 1.02 nm$^{-2}$ (Ref. [48]) | 2D |